\documentclass[manuscript,screen]{acmart}

\AtBeginDocument{%
  \providecommand\BibTeX{{%
    \normalfont B\kern-0.5em{\scshape i\kern-0.25em b}\kern-0.8em\TeX}}}

\copyrightyear{2021}
\acmYear{2021}
\setcopyright{acmcopyright}\acmConference[CHI '21]{CHI Conference on Human Factors in Computing Systems}{May 8--13, 2021}{Yokohama, Japan}
\acmBooktitle{CHI Conference on Human Factors in Computing Systems (CHI '21), May 8--13, 2021, Yokohama, Japan}
\acmPrice{15.00}
\acmDOI{10.1145/3411764.3445494}
\acmISBN{978-1-4503-8096-6/21/05}

\begin{document}

\title[Exploring Asymmetric Roles in Mixed-Ability Gaming]{Exploring Asymmetric Roles in Mixed-Ability Gaming}

\author{David Gonçalves}
\affiliation{%
 \institution{LASIGE, Faculdade de Ciências, Universidade de Lisboa}
 \city{Lisboa}
 \country{Portugal}}
\email{fc52438@alunos.fc.ul.pt}

\author{André Rodrigues}
\affiliation{%
 \institution{LASIGE, Faculdade de Ciências, Universidade de Lisboa}
 \city{Lisboa}
 \country{Portugal}}
\email{afrodrigues@fc.ul.pt}

\author{Mike L. Richardson}
\affiliation{%
 \institution{Department of Psychology, University of Bath}
 \city{Bath}
 \country{United Kingdom}}
\email{mr945@bath.ac.uk}

\author{Alexandra A. de Sousa}
\affiliation{%
 \institution{Centre for Health and Cognition, Bath Spa University}
 \city{Bath}
 \country{United Kingdom}}
\email{a.desousa@bathspa.ac.uk}

\author{Michael J. Proulx}
\affiliation{%
 \institution{Department of Psychology, University of Bath}
 \city{Bath}
 \country{United Kingdom}}
\email{mjp51@bath.ac.uk}

\author{Tiago Guerreiro}
\affiliation{%
 \institution{LASIGE, Faculdade de Ciências, Universidade de Lisboa}
 \city{Lisboa}
 \country{Portugal}}
\email{tjvg@di.fc.ul.pt}

\renewcommand{\shortauthors}{D. Gonçalves, A. Rodrigues, M. L. Richardson, A. A. Sousa, M. J. Proulx, T. Guerreiro}
 
\begin{abstract}
   The landscape of digital games is segregated by player ability. For example, sighted players have a multitude of highly visual games at their disposal, while blind players may choose from a variety of audio games. Attempts at improving cross-ability access to any of those are often limited in the experience they provide, or disregard multiplayer experiences. We explore ability-based asymmetric roles as a design approach to create engaging and challenging mixed-ability play. Our team designed and developed two collaborative testbed games exploring asymmetric interdependent roles. In a remote study with 13 mixed-visual-ability pairs we assessed how roles affected perceptions of engagement, competence, and autonomy, using a mixed-methods approach. The games provided an engaging and challenging experience, in which differences in visual ability were not limiting. Our results underline how experiences unequal by design can give rise to an equitable joint experience.
\end{abstract}



\begin{teaserfigure}
  \includegraphics[width=\textwidth]{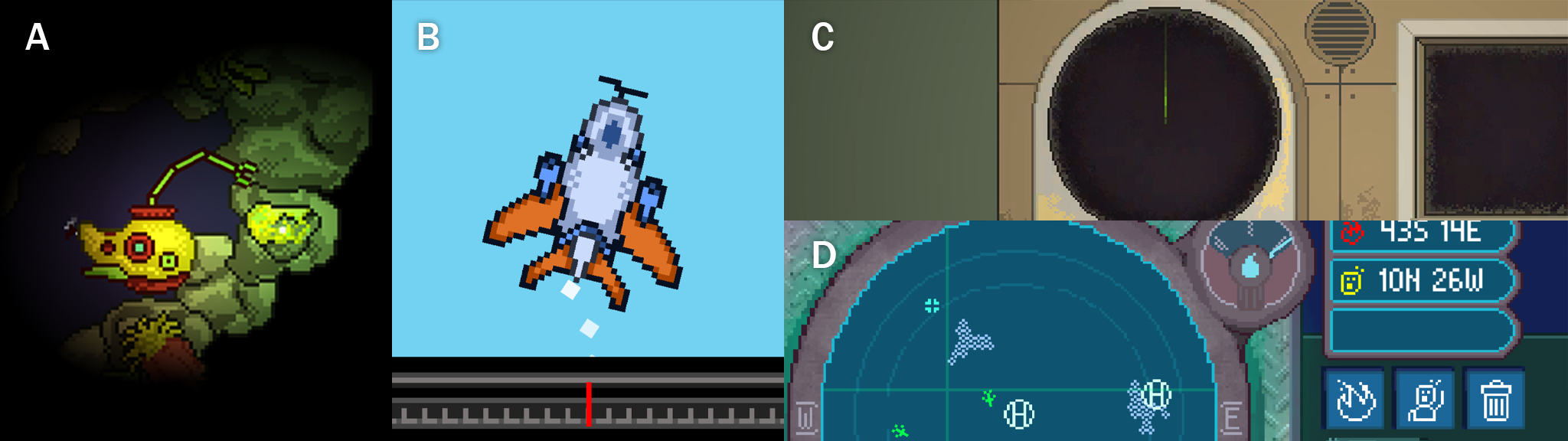}
  \caption{Screenshot details presenting the roles in the games: Pilot (A) and Engineer (C) in Rescue: Under Pressure. Pilot (B) and Engineer (D) in Rescue: Mayday}
  \Description{Sections A, B, C and D represent the pixel-art graphics when playing each of the four different game roles. Section A (Pilot in SUB) shows a submarine, seen from the side, close to a wall of rocks. A mechanical claw protruding from the submarine's hatch is aiming towards a colored rock which is actually a mineral. The central area is lit by a spotlight around the submarine and everything outside of that is black. Section B (Pilot in AIR) shows an airplane viewed from above, on the sky. Particles are emitted from the tail of the plane, which suggest it is moving. A strip at the bottom represents the available and selected radio frequency. Section C (Engineer in SUB) shows a system with a speaker and two monitors, one of them resembling a sonar/radar display. Section D (Engineer in AIR) shows a system with several buttons and monitors. On a large monitor, colored shapes are marked, among which helipad symbols, and minimal representations of clouds and airplanes are discerned. Each smaller monitor shows coordinates of a rescue task and a symbol that indicates whether it is a fire or a person in danger. One of these small monitors has nothing typed and it would remain empty until the Pilot answered a new distress call. Below these small monitors, there are three buttons. The first button has the symbol of a flame, the second of a human face and the third of a trash bin. There's also a small circular display with a blue pointer, which is actually the water tank meter.}
  \label{fig:game}
\end{teaserfigure}

\maketitle

\section{Introduction}

Games usually target a stereotypical player, with a defined set of abilities. For example, most game designs assume players can press various keys at the same time and have the ability to process visual stimuli. This makes most games inaccessible for people who have sensory, motor and/or cognitive disabilities. 

People with disabilities are used to playing games that are exclusively designed for them \cite{andrade_playing_2019, yuan_game_2011, archambault_computer_2007}. Similar to mainstream games, the specific targeting does not consider the interplay between abilities and/or preferences. The lack of intersection leads to isolated communities based on abilities \cite{goncalves_playing_2020}. For example, audio games\footnote{ \url{https://audiogames.net/} (Last visited on August 26th, 2020)} are almost exclusively played by people with visual impairments. These games are typically designed disregarding the stereotypical needs of sighted people (i.e. most audio games don’t have visuals or are not considered appealing).

Despite the availability of a wide variety of accessible games, mixed-ability gaming contexts remain a challenge. Either games cannot be independently played by people with disabilities \cite{yuan_game_2011, porter_empirical_2013}, or games that account for their abilities do not generate enough interest from others with different sets of abilities \cite{goncalves_playing_2020, andrade_playing_2019}. In gaming, we are at a point where it is important to find ways to ensure inclusive experiences, ways that envision mixed-ability interaction as an opportunity, rather than a limitation \cite{esenkaya_multisensory}. In particular, ways to design equally engaging and challenging gaming experiences that overcome differences in players' abilities, stigma and promote social inclusion \cite{holloway_disability_2019}. In our work we explored the design of asymmetric roles in pursuing such experiences.

Asymmetry, in a multiplayer gaming context, means players are presented with distinct gameplays, which can differ at different levels (e.g. character abilities, goals, information) \cite{game_design_workshop}. Research has explored asymmetry before and it has been shown to improve connectedness, especially when a tight cooperation is involved through interdependent roles \cite{harris_asymmetry_2019}. 

We explored the design of asymmetric roles with ability-based challenges entwined in an interdependent collaboration as a strategy to cater for different abilities. We focused on mixed-visual-ability scenarios, thus one role did not require vision and was designed around demanding auditory challenges, while the other focused on visual tasks. Past research has found how often games that strive to be accessible, do so by simplifying games, leading players to quickly lose interest \cite{andrade_playing_2019}. Furthermore, the feeling of unfairness is typically a barrier in mixed-visual-ability gaming experiences \cite{goncalves_playing_2020}. As such, a major point in our approach was to ensure players are challenged on par. We designed two games following this approach and conducted an exploratory user study with mixed-visual-ability pairs. We had the following research questions.

\begin{itemize}
    \item Can asymmetric ability-based roles create engaging experiences for both players?
    \item How do interdependent ability-based challenges affect players perceived competence and autonomy?
\end{itemize}

Our results show how ability-based experiences designed to be unequal can result in equitable gaming spaces where differences in abilities are not limiting. Moreover, the approach showed potential in raising awareness about different abilities and overcoming stigmas. We aim to encourage new ways of designing for mixed-ability play and inspire other researchers and game designers on the path to inclusive gaming.

\section{Related Work}

Here we give an overview of accessible games for people with visual impairments, with a special focus on audio games. Next, we inspect mixed-ability play and asymmetry in gaming by presenting example works from industry and research.

\subsection{Accessible games to people with visual impairments}

In recent years, the gaming industry has made significant efforts towards more accessible gaming. Notably, the 2015 PS4 update included many accessibility features such as text-to-speech and resizable fonts. Microsoft debuted Xbox Copilot\footnote{Copilot on Xbox One. \url{https://beta.support.xbox.com/help/account-profile/accessibility/copilot} (Last visited on September 16th, 2020)} in 2017, an ease of access feature that links two controllers to act as one, enabling co-play and player assistance. Also, following the release of the Xbox Adaptive controller in 2018, specially designed for players with motor impairments, Microsoft patented the Xbox Braille Controller \cite{xbox_braille} in 2019, a controller with a built-in Braille display. Game developer company \textit{Ubisoft Entertainment SA} provides audio descriptions for the trailers of its most recent released games \cite{mikel_reparaz_ubisoft_2020}.

However, the adoption of accessibility guidelines is not consistent in mainstream gaming \cite{porter_empirical_2013}. Importantly, most games still remain unplayable by people with severe visual impairments, since they do not benefit from visual enhancements (i.e. color and contrast options) and depend on audio feedback. Past research has explored different strategies to adapt games that require vision to an accessible audio-based gameplay \cite{yuan_blind_2008, westin_terraformers_2004, smith_rad:_2018, atkinson_mainstream_accessible:2006, Kirke2018_musicraft, vi-bowlling, morelli_vi-tennis:_2010}. In most cases a substantial redesign is needed to ensure that the player is able to perceive the information, determine the correct action, and provide input. 

Some games are shown to be successful in this adaptation. \textit{Blackbox} \cite{blackbox} is an awarded iOS puzzle game, where all puzzles were rethought to be accessible through sound and haptics after its initial release. \textit{The Last of Us Part II} \cite{tlou}, a 2020 top-selling game by Naughty Dog, offers more than 60 accessibility options to players \cite{tlou2}. Notably, the game includes an audio-based reliable rendition of the gameplay that allows someone to play with no visuals at all \cite{tlou2}.

Still, \textit{a posteriori} adaptation has limitations \cite{ua-games, interfaces4all}. One of the potential issues is how the core gameplay may significantly change \cite{archambault_computer_2007}, namely when it comes to games that are inherently visual. For example, gameplay in \textit{The Witness} \cite{witness} involves vision-dependent puzzles which would not be convertible to audio interaction without significantly changing the game mechanics. In addition, when it comes to multiplayer experiences, this adaptation brings even more reservations. New strategies are needed to ensure that players with different abilities are able to play together, and that the experience is engaging, challenging and fair for all. 

People with severe visual impairments are often limited to games that are designed from the ground up to be played by blind people (audio games) or games that are compatible with the technologies they use to access devices (text-based games). These games are not, for the most part, popular among sighted people and therefore are also exclusive in a different way \cite{goncalves_playing_2020, andrade_playing_2019}. However, they offer a substantial variety of games that people with visual impairments can play \cite{andrade_playing_2019, urbanek_audiogames:_2019}.

\subsubsection{Audio games}

Audio games are based on audio-based mechanics and interfaces, relying in the conveyance of speech, audio cues and sonification techniques like earcons, sonar and auditory icons \cite{yuan_game_2011}. They can be played without visuals and most of them do not have visuals at all. Notably, the web site \textit{audiogames.net} presents a list of games that grows frequently with new installments and updated versions. The evolution of audio games and the implications for design have been probed before \cite{urbanek_audiogames:_2019, urbanek_unpacking_2019, archambault_computer_2007, yuan_game_2011, study_accessibility_games:khaliq, games_control_pvi_andrade:_2020}, highlighting the variety of game mechanics and their overall popularity among visually impaired gamers.

Andrade et al. \cite{andrade_playing_2019} established the ability to affect the game narrative is a critical factor that attracted blind people to gaming. In fact, many audio and text-based games have a narrative focus. Some also feature rich complex worlds and give players the opportunity to explore them. Previous work has inspected alternative navigation systems in games, namely audio \cite{atkinson_mainstream_accessible:2006, matsuo_shadowrine:_2016, sanchez_navigation_children:2010, trewin_powerup:_2008, trewin_virtual_worlds:2008,westin_terraformers_2004, white_3d_virtual:2008}, text-based \cite{folmer_textsl:2009} and haptic \cite{matsuo_shadowrine:_2016, sepchat_tactile_games:2006, sanchez_navigation_children:2010, white_3d_virtual:2008} interfaces. For instance, in \textit{AudioQuake} \cite{atkinson_mainstream_accessible:2006}, an adaptation of the popular \textit{Quake} \cite{quake}, researchers use earcons to facilitate in-game interaction. In some audio-based games there’s an analogy with real-life navigation systems, such as compass and sonar like features \cite{games_control_pvi_andrade:_2020, westin_terraformers_2004}. \textit{Musicraft} \cite{Kirke2018_musicraft} is an audio game inspired by Minecraft, in which the world is depicted by abstract musical representations.

In contrast to narrative and exploration focused games, action audio games are typically demanding in terms of dexterity, implying coordination and reaction challenges. Yuan and Folmer \cite{yuan_blind_2008} explored the adaptation of a popular rhythm game, \textit{Guitar Hero} which, despite being a game about musical performance, depends on visual interaction. Adaptation leveraged the use of a glove with small pager motors attached to the tip of each finger, which transmit the information of the notes that must be played. The adaptation of exergames for visually impaired people is also explored in previous work \cite{morelli_vi-tennis:_2010, vi-bowlling}.

\subsection{Mixed-ability gaming}

There are various initiatives to create games accessible for people with visual impairments. Likewise, others that seek to develop accessible games for those with motor \cite{gerling_last_2014, powerball, hobbs_gameon_2012}, hearing \cite{Mascio2013DesigningGF} and cognitive \cite{ohring_webbased_autism} disabilities. Previous work surveyed some of these efforts \cite{yuan_game_2011}. However, research and industry tend to depict accessibility as a way of allowing everyone to experience games, but they typically don't consider scenarios in which people experience them together.

Previously, Gonçalves et al. \cite{goncalves_playing_2020} approached people with and without visual impairments, seeking to characterize mixed-visual-ability gaming experiences. The work outlined two main factors: first, the habits and gaming experience of people with different visual abilities typically do not intersect, especially at the extremes of visual ability (i.e. sighted people play AAA video games, blind people play audio games); second, although these groups find ways to adapt the games to different needs, gameplay is not designed to fit these needs and, therefore, the experience is usually unbalanced and less engaging and/or unfair for someone. 

It is also very difficult to adapt certain visual challenges to audio-only challenges, without losing information or control over the gameplay \cite{smith_rad:_2018}. Smith and Nayar \cite{smith_rad:_2018} developed the \textit{RAD}, an auditory display for gamers with visual impairments aiming for equal accessibility in racing games. Grabski et al. designed \textit{Kinaptic} \cite{grabski_kinaptic}, a tag-like game to be played by a blind and a sighted person. The sighted player interacts through a Kinect and TV, while the blind player relies on a haptic device, wind simulation and 3D sound. The asymmetric setup and multimodality proved to be important to ensure players had a fair winning chance. Inspired by the competitiveness of party games, Imbriani et al. \cite{imbriani_wata_2018} developed the fighting game \textit{WaTa Fight}. Playtesting with mixed-visual-ability groups has shown the game stimulated sociality and the sense of integration.

Jeremy Kaldobsky is an active audio game developer\footnote{Aprone's Accessible Software and Games. \url{https://www.kaldobsky.com/ssl/audiogames.php} (Last visited on September 16th, 2020)}, whose most games feature minimalist graphics. One example is Swamp which is referred to by blind gamers as one of recent games played with sighted friends and family \cite{andrade_playing_2019}. Hicks et al. \cite{hicks_juicy_2019} have shown that graphics, even when purely functioning as an embellishing element conveying redundant information, improve player engagement. Similarly, past work suggests that appealing graphics in audio games are an important factor to engage sighted players \cite{goncalves_playing_2020}. 

Previous work has also explored game design for groups with different motor abilities \cite{powerball, gerling_last_2014}. In \textit{Last Tank Rolling} \cite{gerling_last_2014}, one wheelchair-user and one able-bodied player collaborate in a war scenario. The first player controls a war tank with the movement of their wheelchair and the second controls a soldier using full body gestures. Graf et al. \cite{graf_igym_2019} designed \textit{iGym}, a competitive air hockey inspired exergame, with ``\textit{adjustable game mechanics}” to allow for mixed-motor-ability play. The authors discuss how the adjustment of parameters, in some cases, ensured fair competition by suppressing the advantage of able-bodied players and, in others, it was perceived as an undesirable measure that only reversed the advantage by over facilitating the challenge of players with motor disabilities.

There is research into Universally Accessible (UA) Gaming \cite{ua-games}. Namely, there's the adaptation of traditional games like chess \cite{grammenos_ua-chess:_nodate} and tic-tac-toe \cite{ossmann_computer_2008} as well as complex games that include free world navigation \cite{trewin_powerup:_2008}. This work is incredibly informative for game accessibility by discerning a diverse set of accessibility design options and integrating them into practice. It proves to be above all very important for the establishment of accessibility guidelines, which are now published and disseminated in various forms \cite{hamilton_guidelines_2012, bors_guidelines_2015, mies_guidelines, gameover_grammenos}. Grammenos et al. \cite{grammenos_access_2006}, following an UA Gaming approach, present the concept of Parallel Game Universes. In this, customized profiles are used to adapt the interface and difficulty to the player. This way, players can play in different 'game universes', which are bonded, in this case within a cooperative experience. However, by proposing the same role and task to both, the solution proposed is to reduce the difficulty of the game for players who cannot keep up with the challenge. We point out the negative effect of this specific approach --- the portrayal of players with disabilities as less capable and their contribution as less valuable.

\subsection{Asymmetry in gaming}

Asymmetry can arise at various levels in gameplay (e.g. players have different character abilities). In some games a complete asymmetry is included within the design (i.e. players have very different ways of interacting and share little or no mechanics) \cite{game_design_workshop}. For example, \textit{Cook, Serve, Delicious!} \cite{csd!} has a co-op mode where one player is cooking and the other player is managing orders. In \textit{Clandestine} \cite{clandestine}, players play two interdependent roles --- the spy, which involves a typical third person stealth gameplay and the hacker, responsible for cracking infrastructure and defeating security systems through a birds-eye view of the map and a grid display. It's heavily reliant on patience, communication and teamwork, since the contribution of the two players is imperative for the secret infiltration success. 

Harris and Hancock \cite{harris_asymmetry_2019} found, through a study with a collaborative two-player game, that social presence and connectedness are higher in asymmetric play than in symmetric play, and even higher when tightly-coupled collaboration (i.e. higher interdependence) is involved. It has previously been argued that assuming an explicit asymmetry in gameplay could be a way to cater for different players and their preferences \cite{goncalves_playing_2020, extra_credits_asymmetry}. Asymmetry has previously been used in research for the design of mixed-ability games, although the focus was not the use of asymmetry --- in aforementioned research \cite{grabski_kinaptic, gerling_last_2014, graf_igym_2019}. Notably, in \textit{Last Tank Rolling} \cite{gerling_last_2014}, the movement of the wheelchair is metaphorically and physically linked with the control of a virtual war tank, which is tougher and more powerful, but slower than the foot soldier controlled by the second player. In this example, each role is designed according to the abilities of each player, turning what is normally portrayed as a limitation (the use of a wheelchair) as an empowering element and a valuable contribution. Also, the emergent interdependence of such setting was a catalyst for social play. To the best of our knowledge, asymmetric roles have not been further explored or evaluated in mixed-ability contexts.

\section{Ability-based Asymmetric Roles in Mixed-Ability Gaming}

Wobbrock et al. \cite{wobbrock_ability-based_2011} introduced the concept of ability-based design as a replacing set of questions to model accessible computing. Ability-based design is grounded on ``what can a person do?", emphasizing that the focus is on abilities, not on limitations. It places the responsibility for adaptation in the system, which should be user-adaptable (universally usable). It's noted that universal approaches often search for ``what can everyone do?”, while ability-based takes it from a design-for-one perspective, that asks ``what can you do?”. Similarly, our work is influenced by the Social Model within disability studies, where the disability is not within one's body, but in the oppressive social environment \cite{marks97}, as games are designed for one stereotypical user \cite{goncalves_playing_2020} disregarding mixed-ability play.

More recently, Bennett et al. \cite{bennett_interdependence_2018} point out the negative side of assuming people's independence as the goal of accessibility, being that everyone is actually interdependent, relying on others for sustenance, community, and care. From this perspective, the use of assistive devices can provide full autonomy to the user, but can also picture users as ``\textit{vulnerable or incapable}". The advantages of acknowledging the essential interdependence existing between all humans are highlighted. Authors argue that an interdependence frame established in personal relationships may emphasize contributions from people with disabilities and defy traditional hierarchies that rank abilities \cite{bennett_interdependence_2018}. 

\subsection{Concept}

From this framing comes the idea of an asymmetric gameplay in which each player performs ability-based tasks, contributing to a common goal --- an interdependent group of players that rely on each other to succeed. In short, instead of looking for a single gameplay that can be experienced by players with different abilities, in this approach we explore the entwining of different gameplays in an interdependent collaboration as a potentially effective design choice to achieve an engaging and challenging game to both players [Table \ref{tab:concept}]. 

\begin{table}[]
  \caption{Concept Table.}
  \label{tab:concept}
  
\begin{tabular}{p{0.22\textwidth} p{0.72\textwidth}}

\multicolumn{1}{p{0.22\textwidth}}{\textbf{Concept}} & {\textbf{Description}} \\ \hline 

\multicolumn{1}{p{0.22\textwidth}}{\textbf{Ability-based}} & Challenges are designed based on people abilities (through visual/auditory challenges). The auditory role is vision-independent --- the task is not easier for a sighted person. The auditory role has complementary graphics (important for low vision) but they do not offer an advantage. Similarly, the visual role may have complementary music and sound effects but they are not relevant to the challenge. \\ \hline

\multicolumn{1}{p{0.22\textwidth}}{\textbf{Complete asymmetry}} & Roles intersect in game dynamics (collaboration, communication), but not mechanics --- mechanics are radically different. \\ \hline

\multicolumn{1}{p{0.12\textwidth}}{\textbf{Collaboration}} & The cooperation of two or more people is necessary, as one person would not be able to perform all tasks on their own. \\ \hline

\multicolumn{1}{p{0.22\textwidth}}{\textbf{Interdependence}} & Players depend on each other to succeed. Avoid the perception that one player is being ``assisted” by another. \\ \hline

\multicolumn{1}{p{0.22\textwidth}}{\textbf{Agency}} & There is decision making on the part of both players. The actions of one influence the other. Avoid the perception that one player is just an ``executor" of the game or another player's orders. \\ \hline

\end{tabular}
\end{table}

\section{Testbed Games}
We designed and developed two games as a proof-of-concept. Below we describe our design process and technical aspects of development. 

\subsection{Design process}

We used the three-tiered model of Hunicke et al. \cite{mda} as a framework to ideate. This model describes how a game is perceived from two different perspectives --- as in designing the game and as in playing the game. These three layers are Mechanics (i.e. operations and actions afforded), Dynamics (i.e. emerged behaviors and strategies) and Aesthetics (i.e. perceived aspects by the player). Our design process starts with the conception of mechanics mapped to abilities --- actions and challenges that depend, for one role on visual abilities and, for the other on auditory abilities. We then proceeded to think about how these mechanics could work in the context of a collaborative dynamic between two players, in which player actions are interdependent. The collaborative dynamics were discussed and iterated, leading to changes in the mechanics associated with each role and the synergies between them. 

The concept of both games comes from picturing ways in which no player has complete information or control over a character, ensuring both game roles are essential to be successful. In the first game, ``\textit{Rescue: Under Pressure}” (\textsc{sub}), the player with visual-based interaction is able to move the character (direct control) and the player with audio-based interaction has a global perception of the scenario (world awareness). Inspired by how sonar-like abilities were used before in audio-based world navigation \cite{westin_terraformers_2004, games_control_pvi_andrade:_2020}, in our first game two players control a submarine, one controlling the sonar (auditory challenge) and the other drives the submarine (visual challenge). 

In the second game, ``\textit{Rescue: Mayday}” (\textsc{air}), we reversed the mapping of roles to abilities. One player pilots a rescue aircraft through audio-based interaction and the second is an air traffic controller, with a strong visual task. We drew inspiration from how airline pilots typically do not fly based on visual information and have to rely instead on the instruments and instructions of air traffic controllers. 

Communication is essential between the two players to achieve success. This is the basis of play and collaboration in both games. We ensured that there is considerable diversity in what each player can do. As such, various ability-based mechanics and synergies were designed to fit this collaborative dynamic. Mechanics imply different skills such as memory, dexterity, which could turn to be inaccessible in other mixed-ability contexts. We could have built the visual role as a mix of visual and auditory challenges (since we did not consider hearing impairments), but we chose to focus the challenges of each role on one sense. We used the word ``Pilot” to name the role of the player who has direct control over the character and ``Engineer” to name the role of the player who has world awareness. For a detailed characterization of each role, see Table \ref{tab:roles}.

\subsection{Development}

We used Unity\footnote{Unity. \url{https://unity.com/} (Last visited on September 16th, 2020)} to develop the two online multiplayer games. All graphics were designed from scratch and sound effects were collected from various free sound libraries and post-edited. All game menus and options are accessible via keyboard with pre-rendered (and post-edited) speech from open-source text-to-speech software. Screen readers do not interact with any game elements (including menus). During development we recruited sighted-sighted (i.e. researchers with screen reader expertise) and mixed-visual-ability playtester pairs, whose valuable feedback was critical to ensure usability. We made available a version for Windows and a version for OSx. A trailer video that presents both games is available\footnote{Games Trailer: \url{https://youtu.be/Sgxxgt-favA}}. Games are also available for download\footnote{Games Repository: \url{https://osf.io/2z3tq/?view_only=8ebb305d04c14552be9b9ed57e3e81f2}}. 

\begin{table}[]
  \caption{Roles and mechanics detailed description}
  \label{tab:roles}
\begin{tabular}{p{0.15\textwidth}| p{0.375\textwidth}|p{0.375\textwidth}}
\cline{2-3}
\textbf{Role / Tasks} & \textbf{Rescue: Under Pressure} & \textbf{Rescue: Mayday} \\ \toprule[0.8pt]
\multicolumn{1}{p{0.15\textwidth}|}{\textbf{PILOT}} & \textbf{Visual challenge} & \textbf{Auditory challenge} \\ \bottomrule[0.8pt]
\multicolumn{1}{p{0.15\textwidth}|}{Moves and acts on the character} & Drives the submarine; Switches between different modes (Collect, Flare, Shoot and Battery Saving). & Pilots the airplane (auditory compass); Lands and takes off; Switches between airplane and helicopter mode. \\ \hline
\multicolumn{1}{p{0.15\textwidth}|}{Identifies near \mbox{elements}} & Spots ores, enemies, blocked passages and the treasure within the light around the submarine; Flares light up the cave. & Audio feedback when the airplane is over rescue spots or under storms; Danger sensor beeps when there’s hazards ahead. \\ \hline
\multicolumn{1}{p{0.15\textwidth}|}{Interacts with those elements} & Operates mechanical claw to catch ores; Shoots torpedoes to incapacitate monsters and open blocked passages. & Opens floodgates to extinguish fires; Maneuvers the helicopter and operates the rope to rescue people. \\ \hline
\multicolumn{1}{p{0.15\textwidth}|}{Supply resources to the Engineer} & Collects ores that are used by the Engineer to charge the battery, upgrades sonar and craft various items. & Answers SOS calls that are redirected to the Engineer (with coordinates and time limit); Regularly reports airplane position. \\ \toprule[0.8pt]
\multicolumn{1}{p{0.15\textwidth}|}{\textbf{ENGINEER}} & \textbf{Auditory challenge} & \textbf{Visual challenge} \\ \bottomrule[0.8pt]
\multicolumn{1}{p{0.15\textwidth}|}{Basic information of the world} & Binaural soundscape gives information on the position and proximity of a monster and treasure (passive sonar). & Map screen with control stations marked; Sees airplane position when the Pilot reports its position. \\ \hline
\multicolumn{1}{p{0.15\textwidth}|}{Actively gathers more information} & Sends pulses that detect ores, monsters and the treasure with information on their position; Upgrades sonar range. & Marks rescue spots on the map according to coordinates of received SOS calls; Sends pulses to detect storms and other airplanes. \\ \hline
\multicolumn{1}{p{0.15\textwidth}|}{Manages various aspects} & Controls battery level (charging it when necessary); Crafts and load items for the Pilot to use (flares and torpedoes). & Watches over the water tank level and the state of the vehicle; Regularly consults the time limits for each task. \\ \hline
\multicolumn{1}{p{0.15\textwidth}|}{Provides verbal guidance to the Pilot} & Communicates a direction (e.g. “North”), ensuring the Pilot is able to find ores and the treasure, and to avoid monsters. & Communicates a direction (e.g. “North”), ensuring the Pilot is able to reach rescue spots, and to avoid storms and other planes. \\ \hline
\end{tabular}
\end{table}

\subsection{Games description}

Both games imply a synchronous interaction in which players have to be efficient to achieve the goals within a time limit. Both games include a tutorial mode and a mission mode and require two players. One player plays as the Pilot and the other as the Engineer. The tutorial guides each player through all available mechanics associated with their role as if they were playing mission mode. The only difference is no time limit to any task, and the prompts to introduce each mechanic gradually. Players have to cooperate to fulfill their tasks. In mission mode, there are mechanics that impose time limits to certain tasks, the objectives, resources and challenges are quasirandomly generated (e.g. in \textsc{sub} ensuring the treasure is at a certain depth). Although the games have a very similar collaborative dynamic, the objectives and the way they manage to achieve them are quite different from game to game.

In \textsc{sub}, two players collaborate to rescue a lost treasure using a submarine. Players lose when battery runs out or if the submarine is devoured by a sea monster. The Pilot drives the submarine and operates various tools [Fig. \ref{fig:game} - A]. The Engineer operates the sonar and uses resources to craft items and upgrades (auditory challenge). They must guide the Pilot, by actively surveying the submarine surroundings using the sonar in the eight cardinal points [Fig. \ref{fig:game} - C].

In \textsc{air}, two players collaborate in air rescue missions with the goal to finish the day without any casualties. Players lose when rescue missions reach a time limit or if aircraft collides with a commercial airplane. The Pilot lands, takes off and pilots the aircraft, operates various tools and regularly reports the aircraft position (20 seconds cooldown) [Fig. \ref{fig:game} - B]. The Engineer controls air traffic, marks the location of rescue operations and must guide the pilot to them [Fig. \ref{fig:game} - D]. A description of role tasks and mechanics is presented in Table \ref{tab:roles} and more detailed information about the two games is also available\footnote{Games Detailed Description: \url{https://osf.io/rktz2/?view_only=720e3f6b788b4e05bb93d6708e7b4e44}}.

\section{Study}

We conducted a remote user study with mixed-visual-ability pairs. Our goal was to understand the potential of the approach in creating inclusive and balanced gaming experiences for players with different visual abilities. Seeking to answer the aforementioned research questions, all pairs played both games. Our research questions focus on understanding how asymmetric ability-based roles impact players' perceptions. To understand the effects of asymmetry irrespective of type of role played, we developed two games to ensure players could experience both. We also expected gaming preferences (i.e. type of game, type of role) to impact perceptions. Thus, it allowed us to situate players' feedback. Participants completed one questionnaire after playing each game successfully (tutorial + mission). Lastly, participants completed a debriefing questionnaire where we encouraged them to reflect upon the experience. The study was approved by the Ethics Committee of our school.

\subsection{Participants}

We recruited 26 participants, 13 pairs, from 4 countries, aged 16-53 (M=32.31; SD=8.54) [Table \ref{tab:participants}]. We made an effort to recruit people with different gaming experiences by publicizing the call in dedicated gaming forums and institutions' social networks. We reached expert gamers, casual gamers and also people who are not used to play [Table \ref{tab:participants}]. People applied to participate in pairs (one person had to reach out and sent both contacts). One pair element was required to be a screen reader user due to visual impairments (B1-B13) and the other sighted (S1-S13). 8 participants were totally blind (i.e. no light perception) and 5 participants (B2, B4, B6, B7, B8) were near totally blind (i.e. visual acuity lower than 20/1000). Participants could not have other kinds of severe impairments (mobility, hearing, cognitive), since games were not designed considering other disabilities.

We purposefully recruited pairs of two non-stranger volunteers (family, friends, etc.) to assess how asymmetry could be leveraged in mixed-ability scenarios where people were already familiar with each other. We chose to evaluate with familiar pairs for two reasons: 1) to enable friends and family to play together (given the lack of opportunities for these experiences to happen, highlighted by previous work \cite{goncalves_playing_2020}), and  2) to ensure participants had a baseline of previous interactions with each other. Also, social closeness, competence and autonomy could be affected by other factors among strangers, and are out of the scope of this work. We compensated all participants who fulfilled the protocol with a €25 (or local currency equivalent) voucher, for their time.

\begin{table}
\centering
  \caption{Demographic information, gaming frequency and participation details of participants (each row represents a pair); Playtime values represent time to complete the tutorial (T) and time playing mission mode (M) and are always rounded down to minutes.\vspace*{-6pt}}
  \label{tab:participants}
  \setlength\tabcolsep{3.8pt}
  \begin{tabular}{cccl|cccl|cc|cc}
  {\small \textbf{ID}}
      & {\small \textbf{GN}}
      & {\small \textbf{Age}}
      & {\small \textbf{Plays}}
      & {\small \textbf{ID}}
      & {\small \textbf{GN}}
      & {\small \textbf{Age}}
      & {\small \textbf{Plays}}
      & {\small \textbf{Relation}}
      & {\small \textbf{Country}}
      & {\small \textbf{\textsc{sub} Playtime}}
      & {\small \textbf{\textsc{air} Playtime}}\\
    \toprule
    \textbf{B1}&F&34&Daily&\textbf{S1}&F&53&Monthly&Friends&Australia&27 (T) 18 (M)&22 (T) 14 (M) \\
    \textbf{B2}&M&28&Monthly&\textbf{S2}&F&26&Occasionally&Family&USA&18 (T) 11 (M)&20 (T) 31 (M) \\
    \textbf{B3}&M&41&Daily&\textbf{S3}&F&16&Occasionally&Family&UK&31 (T) 21 (M) &17 (T) 31 (M) \\
    \textbf{B4}&M&38&Monthly&\textbf{S4}&M&38&Occasionally&Friends&Portugal&24 (T) 17 (M)&15 (T) 18 (M) \\
    \textbf{B5}&F&16&Daily&\textbf{S5}&M&18&Weekly&Family&USA&15 (T) 38 (M)&16 (T) 15 (M) \\
    \textbf{B6}&M&32&Occasionally&\textbf{S6}&F&27&Occasionally&Partners&Portugal&21 (T) 18 (M)&21 (T) 29 (M) \\
    \textbf{B7}&F&40&Never&\textbf{S7}&F&32&Weekly&Friends&Portugal&24 (T) 68 (M)&44 (T) 21 (M) \\
    \textbf{B8}&M&38&Occasionally&\textbf{S8}&F&36&Occasionally&Family&Portugal&21 (T) 20 (M)&41 (T) 19 (M)\\
    \textbf{B9}&M&33&Weekly&\textbf{S9}&M&31&Monthly&Friends&Portugal&27 (T) 53 (M)&23 (T) 35 (M) \\
    \textbf{B10}&F&41&Never&\textbf{S10}&F&38&Occasionally&Friends&Portugal&38 (T) 23 (M)&48 (T) 27 (M) \\
    \textbf{B11}&M&25&Monthly&\textbf{S11}&M&26&Weekly&Friends&Portugal&23 (T) 11 (M)&26 (T) 19 (M) \\
    \textbf{B12}&M&37&Never&\textbf{S12}&F&37&Daily&Family&Portugal&33 (T) 23 (M)&51 (T) 12 (M) \\
    \textbf{B13}&F&35&Occasionally&\textbf{S13}&F&24&Occasionally&Family&Portugal&48 (T) 32 (M)&58 (T) 30 (M) \\
    \bottomrule
  \end{tabular}
  \vspace*{7mm}
  \vspace{-25pt}
\end{table}

\subsection{Procedure}

After filling out the participant form, pairs were contacted via email with instructions and the link to the first game they were required to play. Games were counterbalanced between pairs. Participants had at least one week with the first game and were asked to complete the tutorial (about 20 minutes), and play the mission mode for another 10 minutes. We suggested recording their game session (screen capture + communication audio), but did not make it mandatory. After meeting the requirements, each participant received an email to fill in the questionnaire about the experience (detailed below). Participants were free to continue playing if they wished to. After at least one week, and after completing the experience questionnaire, participants were contacted and given access to the second game (same protocol as described above). After complying with the terms regarding both games, participants had to complete one last online questionnaire about their final thoughts on the approach. All play sessions were carefully analyzed, searching for unexpected difficulties (e.g., quitting the game in the middle of the tutorial). When it happened, we contacted the participants and asked if they had any problems playing, offering our help. Participants were also encouraged to reach out to us if they had any issue or question.

\subsection{Online Questionnaires}

Participants completed four online questionnaires: demographics and gaming habits (Q1), game session (Q2 and Q3) and debriefing (Q4). All questionnaires were built in Microsoft Forms and tested for accessibility.

Q2 and Q3 consisted of the Ubisoft Player Experience Questionnaire (UPEQ) \cite{upeq} and open ended questions. UPEQ comprises subscales of Autonomy, Competence and Relatedness and its 21 items are measured on a 5-point Likert scale, where higher is better. We adapted three items to instead of referring to ``characters" to refer to either ``submarine” or ``aircraft”, according to each game. Additionally, we removed the question \textit{``Other players are friendly towards me"} in the relatedness's subscale, as in the context of our study players were not playing with strangers. The open questions asked the participants about their experience and thoughts on the roles. Q4 consisted mainly of open-ended questions that probed participants' final thoughts regarding the games and the concept. Questionnaires are available\footnote{Study Online Questionnaires: \url{https://osf.io/86sye/?view_only=31311e7aff1c47f295b89a5f0dc340e8}}.

\subsection{Data analysis}

We conducted an exploratory study aiming to uncover possible relationships between the concepts of interest without pre-established assumptions or hypotheses. We performed a mixed deductive and inductive thematic analysis over all open-ended questions of the survey, undertaken in line with Braun and Clarke's suggested strategy \cite{braun_2006}. We first familiarized with the data, by iteratively reading the answers. Even before we collected the responses from all participants, we started to annotate relevant phrases and recurring ideas in the text. We created an initial set of codes deductively informed by our readings and enriched with the concepts that stem from our research questions (e.g. asymmetry, competence, autonomy). Codes were discussed among the authors, revised and added, as more responses were submitted in multiple sessions. In these sessions, using Zoom shared virtual white board, themes were discussed, codes were aggregated in a shared placed, relationships were identified and discussed with examples (quotes) spread throughout the board that illustrated the themes that led the research team discussions. Three authors were involved in the coding process. The final codebook and outline of the themes are available\footnote{Codebook \& Themes: \url{https://osf.io/ydurt/?view_only=bfd44c100f0e4870ab65ee8546f4e0ff}}. 

We used UPEQ to quantitatively measure participants engagement, and analysed its subscales of autonomy and competence (``\textit{each subscale of UPEQ independently predicts measures of engagement in game and are a reliable alternative for direct rating of player experience"} \cite{upeq}). We performed mixed within-between analysis of variance to assess the effects of \textit{game} and \textit{user group} on UPEQ scores: all reported analysis meet the assumptions of normality, sphericity, homogeneity of variances, and equality of covariance matrices.
Two pairs recorded and sent us videos of their sessions, which we reviewed to situate their feedback and illustrate our results. 

We chose to assess engagement due to its relationship with players enjoyment and fun, autonomy as it is intrinsically related with the design concept of interdependence, and lastly, competence due to its relationship with asymmetric roles and the past work findings \cite{andrade_playing_2019, goncalves_playing_2020, smith_rad:_2018} around simplistic or unfair games created when striving for accessibility. Nonetheless, the study does not focus on the quantitative measures collected, rather the qualitative data collected is supported by them.

Using a mixed-method approach, below, we characterize participants' experiences and perspectives on ability-based asymmetric game roles. This characterization is primarily based on the thematic analysis of open ended questions. It is additionally confronted and reinforced by the quantitative data resulting from the administration of UPEQ, game session logs, video recordings, demographics and gaming habits.

\section{Findings}

In \textsc{sub}, players took an average of 27.4 minutes (SD=1.7) to complete the tutorial and played mission mode for an average of 28.1 minutes (SD=3.2) (over multiple attempts). In 31 attempts, pairs were devoured by a monster and in 14 they ran out of battery. One pair succeeded in the mission. In \textsc{air}, players averaged 31.5 minutes (SD=2.9) to complete the tutorial and played mission mode for an average of 23.6 minutes (SD=1.5). None of the pairs was able to successfully complete a mission in \textsc{air}, 43 attempts collided with another plane, and 25 were due to the time limits of the rescue operations. 6 played remotely and used audio call to communicate, 6 played co-located and were close enough to communicate, and one pair (B5 and S5) played co-located via audio call. Individual information about each participant's game session is shown in Table \ref{tab:participants}, namely time playing \textsc{sub} and \textsc{air}.

\subsection{Enabling mixed-ability digital play}

Only two sighted participants (S2 and S13) had previously played games with someone with visual impairments (both had played tabletop games with their study partners). Most visually impaired participants reported previous experiences but pointed out that most games do not meet the different needs of groups with mixed visual abilities. 

\vspace{2pt}
\begin{quote} 
\centering 
``\textit{There aren't many games that both blind and sighted players want to play and can play together. Most games are not accessible and sighted people aren't interested in playing poorly designed blindy games}" -- B1
\end{quote}
\vspace{2pt}

The lack of options for group play among people with different visual abilities was highlighted, which is in line with previous work \cite{goncalves_playing_2020}. The games in the study were portrayed by participants as an exception to the rule. Participants expressed a sense of gratification for being able to play digital games with friends (``\textit{we shared an adventure without limitations}``-- S9) and family. 

\vspace{2pt}
\begin{quote} 
\centering 
``\textit{I haven’t been able to find any games that are fun for my brother and I to play together. [...] I feel like it’s a great concept and it was a lot of fun being able to play with my brother.}" -- B5
\end{quote}
\vspace{2pt}

Participants believed the approach provided an inclusive experience and attributed its cause to the asymmetry of feedback and tasks. Referencing asymmetry as: ``\textit{[...] fundamental aspect to allow inclusive fun.}" (B4); ``\textit{[...] focused on the best capabilities of each one, games become rewarding for both players}" (S4). One participant highlighted how mixed-ability gaming with balanced roles, as an inclusive experience, can be leveraged for bonding activities.

\vspace{2pt}
\begin{quote} 
\centering 
``\textit{[Mixed-visual-ability gaming] inspires bonding between people, trust, team spirit. [...] If there are games with roles for people with visual impairments this can help this integration and equality between partners. I really liked the idea, mainly because the person with visual impairments was not given an easier or less important role, but a role as important as that of the other player.}" -- B7
\end{quote}


\subsection{Engagement}

\begin{table}
\centering
  \caption{Quantitative data resulting from the administration of UPEQ. All values are presented in mean (standard deviation) pairs, for \textsc{sub} and \textsc{air}, given all participants (A), just visually impaired participants (VI) and just sighted participants (S).  \vspace*{-6pt}}
  \label{tab:upeq}
  \begin{tabular}{l|lll|lll}
  {}
      & {\small \textbf{\textsc{sub}-A}}
      & {\small \textsc{sub}-VI}
      & {\small \textsc{sub}-S}
      & {\small \textbf{\textsc{air}-A}}
      & {\small \textsc{air}-VI}
      & {\small \textsc{air}-S}\\
    \toprule
    \textbf{Average score}&3.95 (0.08)&3.89 (0.40)&4.02 (0.36)&3.65 (0.11)&3.68 (0.60)&3.63 (0.59) \\
    \textbf{Competence score}&3.67 (0.46)&3.65 (0.44)&3.68 (0.49)&3.09 (0.98)&3.22 (1.10)&2.97 (0.89) \\
    \textbf{Autonomy score}&3.83 (0.56)&3.76 (0.59)&3.89 (0.54)&3.44 (0.68)&3.41 (0.60)&3.47 (0.70) \\
    \bottomrule
  \end{tabular}
\end{table}

All participants reported a positive experience with at least one of the games. Regarding overall engagement with \textsc{sub}, participants reported an average UPEQ score of 3.95 (SD=.08), (visually impaired M=3.89, SD=.48 and sighted M=4.02, SD=.36). In \textsc{air}, participants reported an average UPEQ M=3.65 (SD=.11), (visually impaired M=3.68, SD=.60 and sighted M=3.63, SD=.59). We conducted a mixed within-between analysis of variance to assess the impact of \textit{game} and \textit{user group} on \textit{Overall Engagement}. There was no significant interaction between \textit{user group} and \textit{game}, Wilk's Lambda=.965, F(1, 24)=8.69, p=.36, partial eta squared=.04. There was a substantial main effect for \textit{game}, Wilks's Lambda=.706, F(1, 24)=9.99, p=.004, partial eta squared=.29, with both groups showing to be more engaged in \textsc{sub}. The main effect comparing \textit{user group} was not significant, F(1, 24)=.046, p=.832, partial eta squared=.002.

When we asked participants to choose the game they preferred, 19 participants answered \textsc{sub} and 5 (one of them sighted) answered \textsc{air}. B10 did not answer the question and B3 chose neither, as he states: ``\textit{I can't really pick as I enjoyed them equally.}"

Overall, the experience with \textsc{sub} was significantly more engaging for both sighted and visually impaired participants. Particularly, sighted participants highlighted how being the Pilot on \textsc{sub} was more engaging. S2 commented it was more intuitive and ``\textit{more fun and easy to move around}". On the other hand, participants with visual impairments highlighted how immersive was being the Pilot of the aircraft:

\vspace{2pt}
\begin{quote} 
\centering 
``\textit{I really liked my role. [...] I really like using the different controls to work on the airplane, going where I wanted it as well. I felt like I was actually doing the mission.}" -- B5, referring to \textsc{air}
\end{quote}
\vspace{2pt}


In both games, most participants enjoyed the challenge of managing several aspects at the same time, when playing as the Engineer. However, some participants perceived the role as limited, admitting they would like to have more agency in the action (e.g. in shooting monsters, in rescuing people). The preferences of the players stood out in their perspectives, suggesting changes in the controls and tasks of their role, and even giving ideas for other games with asymmetric roles. Particularly, the narrative setting of each game influenced the feelings of each player, both positively (``\textit{I really like the sea. [..] I really imagined the underwater scenario}" -- B7), and negatively.

\vspace{2pt}
\begin{quote} 
\centering 
``\textit{Maybe, airplanes, storms and fires are not something I find particularly nice. [...] Apart from the setting, it was a nice game to play with a blind friend.}" -- S2, referring to \textsc{air}
\end{quote}
\vspace{2pt}

B5 mentioned she tried to play text-based games with her brother, but pointed out that, in comparison with these, the games in the study ``\textit{were fun because you were actually moving around and doing something}". Audio-based mechanics were praised by participants with visual impairments, such as the helicopter mode (``\textit{I want more of that}" -- B2) and active sonar (``\textit{Really interesting concept using sonar to find different types of materials under water.}" -- B5).

\subsection{Competence \& Autonomy}

Regarding \textsc{sub}, participants reported an average competence of 3.67 (SD=.46) (visually impaired M=3.65, SD=.44 and sighted M=3.68, SD=.49). In \textsc{air}, participants reported an average competence M=3.09 (SD=.98), (visually impaired M=3.22, SD=1.1 and sighted M=2.97, SD=.89).

We conducted a mixed within-between analysis of variance to assess the impact of \textit{game} and \textit{user group} on \textit{Competence}. There was no significant interaction between \textit{user group} and \textit{game}, Wilk's Lambda=.979, F(1, 24)=.524, p=.476, partial eta squared=.02. There was a substantial main effect for \textit{game}, Wilks's Lambda=.719, F(1, 24)=9.36, p=.005, partial eta squared=.28, with both groups showing to feel more competent in \textsc{sub}. The main effect comparing \textit{user group} was not significant, F(1, 24)=.204, p=.656, partial eta squared=.008.

Regarding \textsc{sub}, participants reported an average autonomy score of 3.83 (SD.56), (visually impaired M=3.76, SD=.59 and sighted M=3.89, SD=.54). In \textsc{air}, participants reported an average autonomy score of 3.44 (SD=.68), (visually impaired M=3.41, SD=.67 and sighted M=3.47, SD=.72). A mixed within-between analysis of variance showed no significant interaction between \textit{user group} and \textit{game} on \textit{Autonomy} (Wilk's Lambda=.996, F(1, 24)=.105, p=.749, partial eta squared=.004). There was a substantial main effect for \textit{game}, Wilks's Lambda=.701, F(1, 24)=10.23, p=.004, partial eta squared=.29, with both groups showing to feel more autonomous in \textsc{sub}. The main effect comparing \textit{user group} was again not significant, F(1, 24)=.224, p=.640, partial eta squared=.009.

Participants pointed out how the games were challenging, with the UPEQ item - \textit{"I was good at playing"} having the lowest average score M=2.63. This suggests participants did not feel they had mastered the game in a single play session. As already mentioned, only one of the pairs (B7 and S7) managed to successfully complete a mission, in \textsc{sub}. They do not play often [Table \ref{tab:participants}], but showed enthusiasm for the game considering the time they played (68 minutes, over 8 attempts) and answers to the survey. In \textsc{air}, two pairs were able to accomplish three rescue tasks (they extinguished two fires, rescued a person and landed to deliver the person), but none managed to reach the end of the mission.

Participants felt significantly more autonomous and competent playing \textsc{sub} than \textsc{air}. Some perceived \textsc{air} as too challenging which detracted from the experience. 

\vspace{2pt}
\begin{quote} 
\centering 
``\textit{Not knowing the plane's position on the map more often induces a very high degree of difficulty that prevents the game from being more engaging.}" -- S4, referring to \textsc{air}
\end{quote}
\vspace{2pt}

In \textsc{air}, the Engineer can only see the aircraft's position on the map when the Pilot reports it. We have limited the reporting feature to once every 20 seconds. Its intermittence coupled with the aircraft continuous movement made navigation time sensitive, thus perceived as more demanding. Some pairs felt frustrated because they were unable to reach the relevant locations to complete the objectives. 

Participants described how they reflected on the experience and identified points where they could improve as a team, figuring out how to communicate efficiently. In a particular case, which we were able to observe in full through video recording, a pair (B4 and S4) found an unexpected way to achieve an understanding of roles and collaborative dynamics in \textsc{air}. S4 was not understanding how the aircraft's navigation worked and was not sure how to give directions, so the pair decided to reverse roles. Although they were unable to complete the tutorial (the role of Engineer is not accessible to B4), it was enough to reach an understanding and to collaborate in harmony. In this case, participants were unable to overcome one of the challenges that the game proposes --- reach reconciliation through effective dialogue --- and made this curious decision.

The Pilot's dependence on the Engineer was highlighted by some. However, others mention, in \textsc{sub}, the Pilot has more freedom to explore and have a reference of where ores can be, by following the walls.

\vspace{2pt}
\begin{quote} 
\centering 
``\textit{I found there is a logical dependence on the Engineer. But there is also the possibility for the Pilot to find an ore [close by]. There is a sense of freedom of action.}" -- S9, referring to \textsc{sub}
\end{quote}
\vspace{2pt}

In fact, unlike the Pilot in \textsc{air}, the Pilot in \textsc{sub} has a visual reference of where important elements may be (an idea of "path", formed by the claustrophobic cave). This led to a sense of autonomy, even though, in both games, the Pilot can only perceive elements at a minimum distance and has no way of avoiding dangers without the help of the Engineer. This perception is explicable given that, although there is equivalent information in \textsc{air} when close to fire and rescue locations, the feedback is more spaced out.

While the approach was effective in acknowledging and suppressing differences in abilities, other differences had an impact on the experience. For example, B2 played first \textsc{sub} and explains how the experience has become unbalanced because his partner ``\textit{was not an experienced gamer, so it was difficult for her to find objects and move around the space"}. Likewise, after playing \textsc{air}, B2 commented ``\textit{there's no way I can help the Engineer be better. So if they are struggling and I'm just going in circles, it's not very useful"}. This happened with other pairs. For example, B1, who plays digital games daily, mentions the role of Engineer in \textsc{air} implied a task in which there was "\textit{too much to concentrate on for her [partner]}". 

This was expected, as players have to be efficient playing their roles to succeed as a team. Some participants recognized they would have preferred to have more autonomy during the game: ``\textit{I wanted to take charge, but my role is not one that allows me to take charge.}" (B2, referring to \textsc{air}). For some pairs, it is likely that a loosely coupled interdependence would be more appropriate. B2 adds that ``\textit{a collaborative game where you're two separate entities, like a wood cutter and a builder, would be better, as then we will be in control of our own section.}".

\subsection{Equity \& Awareness}

The collaboration through game roles with tasks and interactions mapped to abilities was associated with social inclusion. With UPEQ, we did not find significant interaction between user group and game in engagement, autonomy or competence. Although experiences with the games were different for all pairs, for most, games were successful in creating a space to have fun, irrespective of abilities.

\vspace{2pt}
\begin{quote} 
\centering 
``\textit{In the end, inclusion and team spirit win, and the barriers between people become more blurred.}" -- B6
\newline
``\textit{On the other side there was just another player with whom I was having fun, and there was no perception that he was a person with a disability.}" -- S4
\end{quote}
\vspace{2pt}

For most participants, control over the collaboration was balanced and tasks were evenly divided through the two roles. Participants mentioned this in relation to both games. The collaborative dynamic was described as a shift in leadership from one role to another depending on what the game status required.

\vspace{2pt}
\begin{quote} 
\centering 
``\textit{I don’t think there was one person who was more in charge. I needed the Pilot to control the submarine, but he needed me to find different objects and monsters [...]. I like that there wasn’t a bigger role.}" -- B5, referring to \textsc{sub}
\end{quote}
\vspace{2pt}

Games offered different gameplays to each player, which, in this case, allowed for a balanced joint experience. Participants' perceptions illustrate how equality does not mean equity (``\textit{[...] giving players a sense of equal circumstances regardless of their characteristics.}" -- B6). Moreover, games showed potential in raising awareness of different abilities, through a social and playful experience. Participants highlighted how they could be leveraged to educate people who may have preconceptions about the impact of a visual disability.

\vspace{2pt}
\begin{quote} 
\centering 
``\textit{It would be a great way of educating sighted people in to what blind people can do in a social non-educational environment.}" -- B3
\end{quote}

\subsection{Information asymmetry}

Participants enjoyed how game mechanisms stimulated communication and interaction between players. This was highlighted in both games, since in both the Pilot depends on clear and succinct indications, not being aware of what surrounds them from a distance. One of the challenges both games featured was the need to figure out how to communicate effectively given the asymmetry of information available for each player. S1 comments how the games ``\textit{allow team work and [are] a great teacher of proper communication between players.}". Information asymmetry was recognized as the main catalyst.

\vspace{2pt}
\begin{quote} 
\centering 
``\textit{The Engineer and I played next to each other. I could see his screen but it didn't give me any clues, which increased the verbal interaction between us. I know the Engineer was notified of the missions being accomplished because he cheered.}" -- S2, referring to \textsc{sub}
\end{quote}
\vspace{2pt}

The tight communication was a highlight of the experience for most participants. However the repetitive nature of their tasks made communication "boring" for some, who suggested alternative forms of communication --- e.g. ``\textit{like a high five emoji or something visible/hearable}" (S2), moving beacons of light, as well as new game mechanics.

\vspace{2pt}
\begin{quote} 
\centering 
``\textit{I feel like maybe more aspects of the game should be us trying to communicate with each other. Like when I drop the rope, he has to guide it or something.}" -- B5, referring to \textsc{air}
\end{quote}
\vspace{2pt}

Information asymmetry caused tensions between players, as it required pairs to solve the problem together and establish a consensus despite neither having all the required information. In the case of S6 and B6, the asymmetric experience caused divisive moments.

\vspace{2pt}
\begin{quote} 
\centering 
``\textit{The interaction was not so good because, in the beginning, we were always seeing who was to blame for not being able to achieve the goal. Me saying I was giving the indications of what I was seeing and he saying he was following the indications he was hearing.}" -- S6, referring to \textsc{air}
\end{quote}
\vspace{2pt}

During design phase, we made decisions to keep information asymmetric, in order to stimulate communication. For example, we discussed whether in \textsc{sub}, the Pilot should have graphical indications of battery level and ammunition. We decided not to, since it is up to the Engineer to control these aspects and to communicate them to the Pilot. Some pairs were successful in finding an efficient way to communicate. As mentioned, one pair was more ingenious and tried to play the tutorial in opposite roles. Participants also highlighted the unity and trust that must exist between the two players to be able to work better as a team.

\vspace{2pt}
\begin{quote} 
\centering 
``\textit{We have to trust our game partner and accept the information he gives us. Therefore, trust, constant interaction and good communication are essential to fulfill the mission.}" -- B7, referring to \textsc{sub}
\end{quote}


\subsection{Inherent Inaccessibility of Ability-Based Roles}

Participants commented how they got to try both roles (Pilot and Engineer) accordingly to the game, and found the ability-based roles to be fundamental for inclusion and equity. However, 3 participants (B1, B2 and B13) mentioned that by having only one accessible role is not very inclusive because the game cannot be played by two visually impaired people. The way we designed the asymmetry fits a very specific scenario: mixed-ability pairs. Some participants mentioned that designing for mixed-ability gaming is ``\textit{inherently discriminatory}" (B2). Two participants with visual impairments pointed out that one of the biggest barriers is not having sighted people to play with. It was suggested that games should allow people with visual impairments to play among themselves or even to play alone: ``\textit{there should be a way in which we can play alone, so we are not always dependent on the availability of another sighted person.}" (B13)

\section{Discussion}

The study was seen by most participants as an opportunity to share a playful moment with a family member or friend, with whom they do not usually play. Seven participants mentioned this was the first experience in which they knowingly played with someone with a different level of vision. Most participants pointed out the lack of games that can be enjoyed by both people with and without visual impairments. In previous work, the lack of intersection of gaming habits and communities in this context was highlighted \cite{goncalves_playing_2020}, which is also emphasized in our results. We collected valuable feedback from participants' perceptions we consider to be informative for research in gaming and accessibility. Some of these observations are related to the nature of collaborative games and asymmetric games, but we focused here on issues particularly relevant for mixed-ability gaming contexts. 

\subsection{Inclusive fun through asymmetry}

Participants found the games to be engaging and inclusive experiences, with some wishing for more content, customization and variety that would enable them to keep playing together. When it comes to mixed-ability gaming, we argue it is not just a question of people being able to play. The ultimate goal, from a hedonic and social perspective, is to design games that people want to play together. By including two stereotypes in the games through a designed asymmetry, the needs of each player were met. We did not seek for universally accessible gameplay. Rather, we focused design on the individual. It successfully provided an intersection space in which both players felt included, challenged and engaged and in which differences in abilities were not limiting. Moreover, we found for most pairs it successfully provided a balanced collaborative gameplay in which both players felt as valuable contributors. Although the experience was not equal for both, it provided equity.

Our findings suggest that, in a context of mixed-visual-ability, ability-based roles can create engaging experiences for both players, which answers our first research question. The concept proved to have potential and it should inspire other researchers and game designers to explore alternative approaches to achieve inclusive fun, even if they appear to suffer some shortcomings towards universal accessibility.

\subsection{Interdependence with ability-based roles}

Participants considered the collaboration in both games balanced, in the sense that none of the players was leading or as if the leadership was constantly shifting between the roles as the situation required. Both roles, in both games, were highly interdependent, to the point that communication was constantly required, which was, for the majority, a highlight of the experience. The asymmetric nature of the information coupled with the interdependence of the roles required participants to: 1) figure out a way to communicate effectively; and 2) trust each other's judgment and information. This led to roles where both players felt valuable, necessary and where the underlying design approach of each role representing an ability did not have any weight to the overall experience, other than guaranteeing an inclusive game. These findings inform how interdependent ability-based roles can impact players' perceived autonomy and competence, which answers our second research question. 

There were only brief tasks that players could engage on their own (e.g. collect a found ore, build a torpedo). The tight dependence on each other led some players (particularly when gaming expertise were diverse in the pair) to report a lack of autonomy and wish for overlapping mechanisms, feedback or a different division of tasks. Harris and Hancock \cite{harris_asymmetry_2019} have faced this issue when analyzing the impact of different degrees of interdependence on players' perceptions. Their results showed that players felt more connected, engaged and interested when interdependence was tighter. However, similar to what we found, they also showed that less skilled pairs preferred looser interdependence. Harris and Hancock present the concept of rhythm of interdependence, in which the level of dependence between players varies in intensity and dynamics. We argue that in mixed-ability contexts using asymmetric ability-based roles, there is an opportunity to explore how to guarantee a challenging experience for all players without sacrificing the interdependence and asymmetry that appears to be at the center of both players feeling necessary, competent and trusted. We believe that, for example, by providing difficulty customization within each role it is possible to accommodate players of different expertise levels and maintain high interdependence.

\subsection{Different-ability awareness}

The educational value of games was recognized in raising awareness of the capabilities and gaming habits of people with different abilities. Participants mentioned the experience would be a good way to show how people with visual impairments are playing digital games. Although the games were strictly designed to be for mixed-ability contexts, participants highlighted how even two sighted players with no accessibility awareness could play the game together and be educated of how blind gamers play. 

In multiplayer games, players typically have an equivalent general perception of the world and the agents in that world (e.g. in a shooter, one player can see others shooting). In games with complete asymmetry, like ours, players have very different information and this perception is divergent. Participants frequently reported the unawareness of the opposite role (i.e. how it was mechanically and interactively), which in some cases aroused curiosity. This is due to the nature of asymmetric gameplay and the fact that participants could not play in reversed roles (one role is inaccessible). Also, to stimulate communication we made decisions in order to maintain asymmetrical information as much as possible. From players' perspectives, we realized the actions and contributions of each person were for the most part implicit, and had to be communicated verbally. This overall unawareness by design, was associated with ``\textit{blurring}" the abilities of each player. In particular, sighted participants mentioned how they wouldn't be able to tell they were playing with a visually impaired player. This provides an opportunity to explore asymmetric games, as a multiplayer entertainment for visually impaired players, where their disabilities do not need to be conveyed or accommodated.

For one pair, asymmetry information gave rise to tension due to the difficulty in finding consensus. However, the same asymmetry inspired the feeling of unity and trust. We find a potential to break stigmas and raise awareness about different abilities. As already mentioned, the individuals who struggled and had to adapt were the ones with less gaming expertise which is expected giving the games were designed to be challenging. We see reflected the potential recognized by Bennett et al. \cite{bennett_interdependence_2018} of acknowledging the interdependence between people --- valuing each person's contributions and destabilizing traditional hierarchies that rank abilities.

\subsection{Tensions of Ability-Based Game Design}

To minimize the tensions that could arise from having a single role accessible, and the possible connotations with it, we developed two games where a different role was accessible in each. In addition, the two different contexts of play enabled us to explore asymmetric ability-based roles where there is a certain equivalence between the same role in different games [Table \ref{tab:roles}]). 

We acknowledge that having roles inaccessible by design can be controversial, particularly when we consider how difficult it is to get someone sighted to play with, due to the isolated gaming communities. During our recruitment we were contacted by 6 more visually impaired people who wished to participate but did not have any sighted partners they could play with. We argue that the isolated communities and lack of mixed-ability play is a symptom of the lack of balanced and engaging opportunities to play which we address with our approach, by design. Previous work cautioned about the negative effects found in other contexts where there is a separation of technologies \cite{separate}. In this study, we found that assuming an explicit asymmetry in gameplay provided an inclusive, engaging and challenging experience for mixed-visual-ability pairs which is not readily available elsewhere. 

Accessibility should not be an \textit{a posteriori} effort to adapt something to the needs of people with disabilities. This seems to be recurrent in gaming accessibility. We argue it should be about finding ways to include these different needs from the beginning of the concept and design process. It is urgent to think of the population as the complex and diverse compound that it is and to find ways to close the gaps that divide different communities, either because of their abilities, their age, or culture. The more work there is, the more faded the perception that games (and more) have to be done for a single stereotypical player without disabilities and the closer we will be to a world where differences exist and they are proudly embraced by design.

\subsection{Limitations}

We conducted a fully remote study with asynchronous communication with participants. As such, we could not control aspects such as the Internet connection of participants, devices they used to play, etc. Although our game prototypes were able to detect disconnections, they were unable to reconnect if the connection was lost for an extended period of time (above 10 seconds). One group, unfortunately, was unable to participate due to frequent connection issues.

Past work has highlighted the lack of opportunities for mixed-ability play, and how audio games are not appealing to sighted players. Thus, although we found our games to be overall engaging, participants have none to limited past experiences playing together. 

It is important to note that these games are prototypes, which are probably not at the level of what participants are used to playing. Although most participants had an engaging experience, it was also associated with tedious moments as the tasks and the way players solved them started to get repetitive.

\section{Conclusion}

Recognizing that mixed-ability digital play remains rare, we sought to explore a different approach to game design that caters for different abilities. We created prototypes of games in which there's a complete asymmetry in the gameplay design based on stereotypical abilities (i.e. visual and auditory). Although we focused on mixed-visual-ability pairs, we believe the approach is promising to any mixed-ability context, where roles are designed based on a chosen set of abilities. We showed the approach was able to provide an inclusive experience unlike the games participants usually play. Our work shows the potential for novel approaches where equity does not necessarily come from equality.

\begin{acks}

We kindly thank all of our participants. This work was supported by FCT through LASIGE Research Unit funding, ref. UIDB/00408/2020 and ref. UIDP/00408/2020, 
Portugal-UK Bilateral Research Fund (PARSUK BRF) through project ``INPLAY: Designing games for inclusive play with the visually impaired and sighted", and the UKRI Centre for the Analysis of Motion, Entertainment Research and Applications (CAMERA 2.0; EP/T022523/1).

\end{acks}

\bibliographystyle{ACM-Reference-Format}
\bibliography{acmart}

\appendix

\end{document}